\documentstyle[12pt]{article}
\begin{document}
\title{\LARGE \bf
       sd-Shell Study with a \\[5pt]
       Multi-Configuration Mixing Approach\\[5pt]
       designed for Large Scale\\[5pt]
       Nuclear Structure Calculations}
\vspace{15pt}
\author{
        E.~Bender, K.W.~Schmid and Amand~Faessler\\[10pt]
        {\it Institut f\"ur Theoretische Physik}\\
        {\it Universit\"at T\"ubingen}\\
        {\it Auf der Morgenstelle 14}\\
        {\it D-72076 T\"ubingen}\\
        {\it Germany}}
\date{}
\maketitle
\thispagestyle{empty}
\vspace{11pt}
\begin{center}
\bf Abstract
\end{center}
A systematic numerical investigation of a recently developed nuclear
structure approach is presented which diagonalizes the Hamiltonian in the
space of the sym\-me\-try-pro\-jec\-ted Hartree-Fock-Bogoliubov (HFB) vacuum
and symmetry-projected quasiparticle excitations with respect to it. The
underlying HFB transformation, which is assumed to be time-reversal and
axially symmetric, is determined by variation after the projection.
The model allows the use of large basis systems. It has been applied to the
calculation of energy spectra of several even-even, odd-odd and odd mass
nuclei in the $sd$ shell with mass numbers reaching from $A\!=\!20$ to 30.
The Chung-Wildenthal interaction has been used. Good agreement with the 
exact shell model diagonalization and a considerable improvement on a 
previous approach, where the HFB transformation was significantly more 
restricted, is obtained.
\pagebreak
\section{Introduction}
\label{introduction}
\setcounter{page}{2}
\setlength{\parskip}{\baselineskip}
Microscopic nuclear structure models try to explain the excitation energies
and other experimental spectroscopic information starting from a realistic 
effective nucleon-nucleon interaction. An example is the interacting shell 
model. It is generally acknowledged to be the most fundamental theory of a
nucleus \cite{BRU77} \cite{MCG80} \cite{BRO88}. But since the number of 
A-nucleon configurations to be diagonalized increases rapidly with the 
number of single particle states, its application is limited to rather small
basis systems, like the $sd$ shell \cite{SCH87b}. Thus for many problems
the configuration spaces have to be truncated drastically. How to select 
a numerically feasible number of physical relevant configurations is the
central question of all nuclear structure models which use the shell model
as the conceptual basis.
\par
One prescription for such a truncation is provided by the mean field approaches
like the Hartree-Fock \cite{HAR28} \cite{FOC30} or the Hartree-Fock-Bogoliubov
(HFB) \cite{BOG58} \cite{RIN80} theory. They use the variational principle to
describe the ground state of a nucleus by one single Slater determinant.
However, usually in this determinant all physical symmetries are broken. Thus
one has to use projection techniques to find the component with the desired 
quantum numbers in order to obtain physical states. Moreover, in order to 
obtain the energetically lowest solution for a given set of quantum numbers,
one has to perform the projection on the good symmetries before the mean field
is determined by variation.
\par
Following this idea, a whole hierarchy of symmetry conserving mean field
theories and their extensions into multi-configuration mixing methods have
been proposed by some of our group a few years ago \cite{SCH84a}. They are
known as the VAMPIR ({\bf V}ariation {\bf A}fter {\bf M}ean field
{\bf P}rojection {\bf I}n {\bf R}ealistic model spaces) and the MONSTER
({\bf MO}del for handling many {\bf N}umber- and {\bf S}pin-projected
{\bf T}wo-quasiparticle {\bf E}xcitations with {\bf R}ealistic interactions
and model spaces) approaches.
\par
The models of the VAMPIR family, like VAMPIR \cite{SCH84c}, EXCITED VAMPIR
\cite{SCH86a}, FED (FEw Determinant) VAMPIR and EXCITED FED VAMPIR
\cite{SCH89b}, can be used to describe the lowest few states of a given
symmetry, irrespective of their particular structure. Each state is obtained
by variational calculations. So, e.g., the VAMPIR model approximates the
lowest state of certain quantum numbers by just a single symmetry-projected HFB
vacuum. The underlying HFB transformation is determined by variation after
the projection on the desired quantum numbers. Correlating configurations
as well as excited states can be obtained by chains of similar variational
calculations.
\par
On the other hand, the MONSTER approaches are constructed for the description
of complete excitation spectra with respect to a one-body transition operator
\cite{SCH87b} \cite{SCH84b}. Thus they are well suited, e.g., for the
description of giant resonances. States which are reached by a one-body
operator are of similar structure as the initial ground or yrast state.
Therefore they can be described by expanding the nuclear wavefunction around
a symmetry-projected reference vacuum, which may be either a usual HFB, or,
e.g., a VAMPIR solution for the ground or yrast state. This is done in the
MONSTER approach. More explicitly, the MONSTER configuration space is built
out of the symmetry-projected vacuum and all symmetry-projected
two-quasiparticle excitations defined on it. The excited states are 
then obtained by diagonalizing the residual interaction between these
configurations.
\par
In all realistic applications up to now, these models have been simplified
by imposing symmetry restrictions on the underlying HFB transformations.
In the first VAMPIR calculations only real, time-reversal invariant and
axially symmetric HFB transformations, which neither mix proton and neutron
states nor states of different parity, were admitted \cite{SCH84c}
\cite{SCH86a}. With this {\sl real} VAMPIR approach, as it is called, only
states in even-even nuclei with even spin and positive parity can be described.
If a MONSTER calculation is based on such a {\sl real} VAMPIR transformation,
states with different symmetries are described entirely by the
two-quasiparticle components of the wavefunction. Some years ago, this VAMPIR
approach has been improved by using HFB transformations, which are essentially
complex and allow proton-neutron and parity mixing \cite{SCH87a}. Only time
reversal and axiality are still imposed on the transformation. This
{\sl complex} VAMPIR approach can describe states of arbitrary spin
parity in even-even and odd-odd nuclei. It considers many more nucleon
correlations \cite{ZHE89}.
\par
Recently the MONSTER approach has been generalized for the use of such 
{\sl complex} VAMPIR transformations \cite{BEN95a}. It was tested for 
two light even-even $sd$ shell nuclei. The results were encouraging. The 
performance of the {\sl complex} MONSTER description for heavier 
even-even nuclei, for odd-odd and odd-mass nuclei, however, remained an 
open question. In the present paper we try to answer this question by 
a systematic investigation.
\par
For this purpose the {\sl complex} MONSTER model was applied to
several even-even, odd-odd and odd mass systems. The calculations were 
performed again in an $sd$ shell basis, in order to be able to compare
with the exact shell model results. The considered nuclei of each type are
covering the range from very light systems to the ones with the biggest
possible numbers of shell model configurations in the $sd$ shell. The results
are compared to those of the older more restricted {\sl real} MONSTER approach,
too. It is demonstrated that the {\sl complex} MONSTER approach approximates
the shell model results very well and shows a clear improvement on the older
{\sl real} approach.
\par
In the next section the main features of the MONSTER on VAMPIR model are
summarized and the consequences of the symmetry restrictions for the
HFB transformation are shortly explained. In section (\ref{application})
the results of the application are presented and discussed. Finally
conclusions and an outlook are given in (\ref{summary}).
\section{The Model}
\label{model}
The detailed formulation of the MONSTER on VAMPIR approach has been given
elsewhere \cite{BEN95a}. Therefore only the main ideas of the model are
sketched here.
\subsection{The VAMPIR model}
\label{vampirmodel}
A finite model space is defined by a $D$-dimensional set of orthonormal
single particle wave functions, which are eigenstates of some spherically
symmetric one-body potential, e.g. the harmonic oscillator. In second
quantization the corresponding creation and annihilation operators are
denoted by $ \{ c_i^{\dagger}, c_k^{\dagger}, \ldots \}_{D} $ and
$ \{ c_i, c_k, \ldots \}_{D} $, respectively. The indices $i$, $k$
summarize the quantum numbers of a state. It is assumed that
the effective many body Hamiltonian appropriate for this model space
is known and can be represented by a sum of only one- and two-body terms 
\begin{equation}
  \label{hamiltonian}
       \hat{H} = \hat{T} + \hat{V} .
\end{equation}
The one-body part $\hat{T}$ contains the matrix elements of the kinetic energy
(or, if an inert core is assumed, the single particle energies) and the
two-body part $\hat{V}$ the matrix elements of the effective interaction.
\par
Starting from the chosen particle basis, the Hartree-Fock-Bogoliubov (HFB)
transformation \cite{RIN80} is used to define the corresponding quasiparticle
basis with its quasiparticle creation and annihilation operators,
$ \{ a_{\alpha}^{\dagger}, a_{\beta}^{\dagger}, \ldots \}_{D} $ and
$ \{ a_{\alpha}, a_{\beta}, \ldots \}_{D} $. $\alpha$,~$\beta$ enumerate the
quasiparticle states. The HFB transformation is the most general linear
transformation which ensures that the Fermion anti-commutation relations are
fulfilled by the quasiparticle operators. In matrix notation it is given by
\begin{equation}
  \label{hfbtransformation}
       \left(
         \begin{array}{c} a^{\dagger}(F) \\ a(F) \end{array}
       \right)
       =
        F \left(
            \begin{array}{c} c^{\dagger} \\ c \end{array}
          \right) 
\end{equation}
with $F$ being a unitary $2D\!\times\! 2D$ dimensional matrix.
The vacuum $|F\rangle$ for a set of quasiparticles obtained by the HFB
transformation $F$ is defined in the usual way by the request that application
of any quasiparticle annihilation operator onto it should yield zero,
\begin{equation}
    \label{vacuum}
         a_{\alpha}(F) |F\rangle = 0 \quad \forall \: \alpha =1,\ldots,D .
\end{equation}
An $n$-quasiparticle state with respect to the vacuum is defined by
\begin{equation}
  \label{n-qpstate}
       | F \{ a^{\dagger} \}_n \rangle =
            \left( \prod _{\alpha = 1} ^n a^{\dagger}_{\alpha}(F) \right)
              | F \rangle  \quad \mbox{for} \quad n=1 ,\ldots ,D .
\end{equation}
\par
The vacuum as well as the $n$-quasiparticle states contain components of many
nucleon configurations with different angular momenta, different angular
momentum $z$-components, both parities and various proton and neutron numbers
\cite{MAN75}. In order to get physical states which are characterized by good
quantum numbers, projection techniques have to be used for selecting the
components with the desired symmetries. The corresponding projection operator
$\hat{\Theta}^{AT_zI^{\pi}}_{MK}$ is a product of projection operators,
\begin{equation}
  \label{symmetryprojector}
       \hat{\Theta} ^{AT_zI^{\pi}}_{MK} \equiv
         \hat{P}(IM;K) \hat{P}(2T_z) \hat{P}(A) \hat{P}(\pi) ,
\end{equation}
where the first projects on good angular momentum $I$
with $z$-component $M$, the next ones on good isospin $z$-component $T_z$,
good mass number $A$ and finally on good parity $\pi$. Explicit expressions
for them can be found in \cite{SCH87a}. All projectors apart from
$\hat{P}(\pi)$ are integral operators.
\par
Physical configurations are obtained by applying the operator
$\hat{\Theta}^{AT_zI^{\pi}}_{MK}$ on the above introduced quasiparticle
configurations. Since any of the $n$-quasiparticle configurations
$| F \{ a^{\dagger} \}_n \rangle $ built on the vacuum $|F\rangle$
can be looked upon at the same time as a vacuum for another HFB transformation
$F^{\prime}$, it is enough to consider only HFB vacua in the following.
Thus, a physical configuration with a good symmetry $S$, where $S$ represents
the quantum numbers $AT_zI^{\pi}$, is given by
\begin{equation}
  \label{projectedvacuum}
       |F;SM\rangle =
       \sum_{K=-I}^{I}
             \hat{\Theta}^S_{MK} |F\rangle f_K^S .
\end{equation}
The sum over all intrinsic angular momentum $z$-components $K$ has to be taken
in order to avoid a dependency of the projected wave function on the
orientation of the intrinsic reference frame \cite{SCH84a}.
\par
In the VAMPIR approach one symmetry-projected vacuum (\ref{projectedvacuum})
is used to describe the energetically lowest state of a given symmetry $S$.
The configuration mixing degrees of freedom $f_K^S$ and the underlying HFB
transformation $F$ are determined by variation
\begin{equation}
  \label{vampirvariation}
       \delta E^S \equiv
         \delta \frac{\langle F;SM | \hat{H} | F;SM \rangle}
                          {\langle F;SM |  F;SM \rangle}
         = 0 .
\end{equation}
This variation leads to three sets of equations which have to be solved
self-con\-sis\-tent\-ly \cite{SCH87a}. One obtains the optimal description of
the considered yrast state which can be achieved by using only a single
determinant. Generally this VAMPIR approach yields already a rather good
approximation to the actual state \cite{ZHE89}.
\par
A similar approach, the FED VAMPIR, uses a linear combination of a few
symmetry-projected vacua to describe one yrast state. Here, the correlating
configurations, which contribute to one state, are searched for by successive
variational calculations \cite{SCH89b}. Furthermore, by introducing
orthogonality constraints, the VAMPIR and the FED VAMPIR procedures can be
easily extended to the description of excited states. These EXCITED VAMPIR
\cite{SCH87a} and EXCITED FED VAMPIR \cite{SCH89b} models can describe states
of any arbitrary structure. But since each state is searched for by variation,
a numerical application is quite time consuming. Thus the VAMPIR models are
especially well suited for the description of the lowest few states of a
given symmetry $S$.
\par
For problems which require a complete excitation spectrum with respect to
some transition operator, it is better to follow another avenue. 
Since here only specific configurations are needed, one can describe them
by expanding the nuclear wave functions around a suitable VAMPIR vacuum.
This is explained in the next section.

\subsection{The MONSTER on VAMPIR approach}
\label{monsteronvampir}
For states with symmetry $S$ the configuration space is chosen to consist
of the symmetry-projected vacuum 
\begin{equation}
  \label{monsterconfigurationspace2a}
       |F;SMK\rangle \equiv \hat{\Theta}^S_{MK} |F\rangle 
\end{equation}
and the symmetry-projected two-quasiparticle states with respect to it
\begin{equation}
  \label{monsterconfigurationspace2b}
       |F\alpha\beta;SMK\rangle
       \equiv
         \hat{\Theta}^S_{MK} a^{\dagger}_{\alpha}(F) a^{\dagger}_{\beta}(F)
         |F\rangle .
\end{equation}
The underlying HFB transformation $F$ is fixed in a preceding VAMPIR
calculation for an yrast state, which not necessarily needs to have just
the same symmetry $S$. In many cases it is sufficient to use the HFB
transformation obtained for the ground state of the considered nucleus.
\par
A general wavefunction for excited states is then given by
\begin{equation}
  \label{monsterwavefunction}
       |\psi_i(F);SM\rangle
       = \left\{ \sum_{K=-I,\ldots ,+I} |F;SMK\rangle g^S_{0K;i}
          + \sum_{{\alpha < \beta} \atop  K=-I,\ldots ,+I}
                |F\alpha\beta ; SMK \rangle g^S_{\alpha\beta K;i} \right\} .
\end{equation}
The expansion coefficients $g^S$ are obtained by diagonalizing the effective
Hamiltonian in the space of the nonorthogonal configurations
(\ref{monsterconfigurationspace2a}, \ref{monsterconfigurationspace2b}).
\par
The above choice of the model space restricts the MONSTER on VAMPIR
approach to excited states with a structure similar to that of the
underlying projected vacuum. This is here desired, since we are interested
in states predominantly populated by a transition from the vacuum
via a one-body operator. Furthermore it should be mentioned
that the MONSTER approach provides an approximate elimination of spurious
center-of-mass excitations, which can be introduced by the configuration
mixing if more than one major oscillator shell is taken as particle basis
(for details see \cite{BEN95a}).
\subsection[Symmetry restrictions]
           {Symmetry restrictions imposed on the \\
            HFB transformation}
\label{restrictions}
If the most general HFB transformations are allowed, symmetry-projected vacua
of type (\ref{projectedvacuum}) can describe arbitrary states in arbitrary 
nuclei \cite{SCH87a}. Up to now, however, this has not been achieved in any
numerical implementation. Instead, for the existing numerical realizations
of the VAMPIR approaches, certain symmetry requirements were imposed on the
underlying HFB transformations.
\par
In the first VAMPIR calculations \cite{SCH84c} axial symmetry and time-reversal
invariance were assumed, parity and proton-neutron mixing were neglected and
only real HFB transformation coefficients were admitted. As a consequence of
these approximations the {\sl real} VAMPIR approach was only suitable for
positive parity states with even angular momenta in doubly even nuclei.
Performing MONSTER calculations on top of such a {\sl real} VAMPIR solution
these limitations are removed. In the two-quasiparticle approximation not
only states with arbitrary spin-parity in doubly even but also in doubly odd
systems become accessible. However, for the description of odd spin and/or
negative parity states in doubly even nuclei one has to use mean fields
obtained for different spin and/or parity values than the considered one. And
for doubly odd systems one even has to rely on mean fields for neighbouring
systems. Considering as configuration space for the MONSTER calculation
all symmetry-projected one-quasiparticle states
\begin{equation}
  \label{1qpmonsterconfigurationspace}
       |F\alpha;SMK\rangle
       \equiv
         \hat{\Theta}^S_{MK} a^{\dagger}_{\alpha}(F) |F\rangle ,
\end{equation}
arbitrary states in odd systems become approachable, too.
Here again, one has to build the one-quasiparticle spectrum on
the mean field determined for a neighbouring doubly even nucleus.
\par
In the more recent implementations of the VAMPIR calculations parity as well
as proton-neutron mixing are taken into account and essentially complex HFB
transformations are admitted. In this {\sl complex} VAMPIR approach only
axial symmetry and time reversal invariance are kept \cite{SCH87a}.
This introduces many more correlations into the projected vacua than
the older {\sl real} calculations. Now states of arbitrary spin-parity
in both doubly even and doubly odd systems are accessible. The {\sl complex}
vacua contain all possible two nucleon couplings, but some
four-, six- and more nucleon couplings are missing \cite{ZHE89}. Only recently
the MONSTER approach could be extended to such {\sl complex} VAMPIR vacua 
\cite{BEN95a}. Here, in addition to doubly even nuclei, also the calculation 
for doubly odd nuclei can be based on a HFB transformation determined for 
just the considered system itself. And the structures which are missing in 
the {\sl complex} VAMPIR vacuum can be
introduced by the two-quasiparticle admixtures. However, for the description
of odd systems one still has to use the one-quasiparticle approximation
like in the older {\sl real} approach. But now aside from the mean
fields of neighbouring even-even nuclei also the ones of neighbouring
odd-odd nuclei can be chosen to construct the one-quasiparticle
states~(\ref{1qpmonsterconfigurationspace}).
\par
Generally, since the underlying HFB transformation is much more general
and the configuration space is much larger, the {\sl complex} MONSTER
is expected to yield a better description of nuclear spectra
than the {\sl real} one.
\section{A systematic application to {\it sd} shell nuclei }
\label{application}
In \cite{BEN95a} we presented as a first test of the new {\sl complex} MONSTER
approach an application to $^{20}$Ne and $^{22}$Ne. Considering the $sd$ shell
as model space, it was shown that the model is able to reproduce the
shell model spectrum of $^{20}$Ne exactly. For $^{22}$Ne it yielded an
excellent approximation to the shell model result. A clear improvement on
the older {\sl real} approach was found. However, the two investigated nuclei
are even-even systems. Moreover, they are quite light~: for both nuclei
the number of shell model states with a certain spin is comparable to
the number of {\sl complex} MONSTER configurations. How does the approach
perform for heavier nuclei, where the number of states is much larger than
the number of {\sl complex} MONSTER configurations ?  Furthermore, which 
kind of agreement is obtained for odd-odd nuclei and for odd mass systems, 
where the simpler one-quasiparticle description is used ? These questions are
adressed in the present investigation.
\par
Again the $1s0d$ shell was chosen as single particle basis in
order to be able to compare with exact shell model configuration mixing
calculations. The single particle energies 
$\epsilon\,(d{\,5/2}) = -4.15\mbox{ MeV}$,
$\epsilon\,(s{1/2}) = -3.28\mbox{ MeV}$, and 
$\epsilon\,(d\,{3/2}) = +0.93\mbox{ MeV}$ for both protons and neutrons
\cite{AJZ86} and the effective two-body residual interaction 
(the mass-dependent version of the Chung and Wildenthal force \cite{WIL83} 
($\hat{V}(A) = \hat{V}(18) \times (18/A)^{\alpha}$) with $\alpha = 1/3$ 
instead of $\alpha = 0.3$) are the same as in our earlier calculations.
This force is generally accepted to be ``the standard'' force for 
the $sd$ model space. The results are compared in addition to those 
obtained by the more restricted {\sl real} MONSTER approach.
\subsection{Even-even nuclei}
\label{eveneven}
As examples for ``heavier'' even-even $sd$-shell nuclei, we calculated the 
spectra of $^{24}_{12}$Mg, $^{26}_{12}$Mg, $^{28}_{14}$Si and $^{30}_{14}$Si 
within the {\sl complex} and the {\sl real} MONSTER approaches.
For each MONSTER calculation the underlying HFB transformation was determined
by a preceding VAMPIR calculation for the $0^+$-ground state. This HFB
transformation F($0^+$) has then been used for states of all spins in
one nucleus.
\par
The energies obtained by the {\sl complex} and the more restricted {\sl real}
MONSTER have been compared to those from an exact shell model
diagonalization. The result of this investigation is summarized in table~1.
The average deviation of the energies of the yrast states with spin $0^+$
to $5^+$ from the shell model energies is presented for each nucleus. The
deviation is given in keV and in percent of the shell model ground
state energy of the nucleus referred to. The exact ground state energy is also
displayed. For comparison the corresponding values for the nuclei $^{20}$Ne
and $^{22}$Ne of our first application are listed here, too. The table
shows furthermore the minimum and maximum number of shell model configurations
per spin for each nucleus and the corresponding spin value. In addition the
number of shell model states per spin averaged over the states with spin $0^+$
to $5^+$ is given. It should be compared to the number of MONSTER
configurations available in the $sd$-shell. As explained in more detail
in \cite{BEN95a} the {\sl complex} MONSTER provides 57 configurations for 
the description of spin $0^+$ states, 151 for $1^+$, 223 for $2^+$, 259 
for $3^+$, 275 for $4^+$, and always 277 for all higher spin states. In the 
more restricted {\sl real} MONSTER approach less configurations per spin
are available~: 21 for the $I=0$ states, 30 for $I=1$, 61 for $I=2$,
56 for $I=3$, always 60 for $I=5,7,9,11,13$, and always 73 for the
$I=4,6,8,10,12,14$ states, respectively.
\par
As expected for both MONSTER approaches the mean deviation of the yrast
state energies from the exact shell model result, averaged for spins $0^+$ 
to $5^+$,  does increase with the number of actually existing shell model
states. For the {\sl complex} MONSTER the largest deviation measured
relatively to the shell model ground state energy occurs for $^{26}$Mg
and amounts to 1.43\% . For $^{20}$Ne the {\sl complex} calculation reproduced
the exact result. In the {\sl real} case the largest average deviation,
2.42 \% , is found for $^{24}$Mg and the smallest, 0.87\% , for $^{22}$Ne.
For all nuclei the {\sl complex} MONSTER approach yields a clear improvement
on the {\sl real} one. The largest energy gain could be achieved in
$^{24}$Mg, where the averaged deviation of the {\sl real} MONSTER is 3.8
times larger than the one of {\sl complex} MONSTER. The smallest amount
of energy was won for $^{26}$Mg, where the average deviation of the {\sl real}
approach is by a factor of 1.3 larger compared to the {\sl complex} one.
For all even-even nuclei the {\sl complex} MONSTER approach yields an
excellent approximation to the exact calculation.
\par
As an example, in fig.~1, the spectra of $^{28}$Si as obtained by the
{\sl complex} and the {\sl real} MONSTER are shown up to $\sim$~14~MeV
above the shell model ground state energy. For comparison the energies
resulting from the shell model diagonalization are plotted up to $\sim$~12~MeV
excitation energy. It can be seen that the {\sl complex} MONSTER reproduces
the shell model spectrum well and shows a clear improvement on the
{\sl real} MONSTER~: the states are considerably more bound and
many more shell model states can be described.
\par
The nucleus $^{28}$Si was investigated in more detail, since it has the
largest number of states of all $sd$-shell nuclei. There exist in total
93710 states. More precisely, there are 3372 $I^{\pi}\!\!=\!0^+$ states,
9216 $I^{\pi}\!\!=\!1^+$, 13562 $I^{\pi}\!\!=\!2^+$, 15385 $I^{\pi}\!\!=\!3^+$,
15089 $I^{\pi}\!\!=\!4^+$, 12876 $I^{\pi}\!\!=\!5^+$ and
9900 $I^{\pi}\!\!=\!6^+$ states. For the higher spins the number of states
per spin continues to decrease; there is just one state with the highest
possible spin $I^{\pi}\!\!=\!14^+$. In fig.~2 the yrast state energies as
obtained by various models of the VAMPIR and MONSTER family are compared
for the spins $0^+$ to $6^+$. The energies from the exact shell model
calculation are given in the leftmost column. Then, from left to right,
the energies calculated by the {\sl complex} MONSTER, the {\sl complex}
VAMPIR, the {\sl real} MONSTER, and the {\sl real} VAMPIR are shown. Note
that the {\sl real} VAMPIR approach can only describe even spin states.
\par
The average deviation of the energies obtained by the
{\sl complex} MONSTER approach from the shell model, now including the
$6^+$ yrast state, is about 1085~keV or 0.79\% of the shell model
ground state energy of 137.495~MeV. The deviation for the even spin states
is on average somewhat smaller, namely 840~keV, than the one for
the odd spin states, which amounts to 1412~keV. This was found also
for the other investigated even-even nuclei, apart from $^{24}$Mg, where
the difference is insignificant. There the average deviation of the odd
spin states was only 51~keV larger than for the even spin states.
{\sl Real} MONSTER gives for $^{28}$Si an average deviation of 1746~keV,
which amounts to 1.27\% of the shell model ground state energy.
\par
As it is to be expected, the {\sl complex} MONSTER yrast states are more
bound than the ones calculated by {\sl complex} VAMPIR, except for the
$0^+$ VAMPIR state, which is stable with respect to any projected
two-quasiparticle excitations of the same symmetry. Obviously the odd spin
states $1^+$, $3^+$ and $5^+$ are not well described by the {\sl complex}
VAMPIR. When the difference between the binding energies obtained by the
two models is so pronounced, it can be inferred that these state are dominated
by the above mentioned ``missing couplings'', which are not accessible within
the {\sl complex} VAMPIR model. This was also found in Ref.~\cite{BEN95a}, 
where the odd spin yrast states of $^{22}$Ne were rather poorly described by
the {\sl complex} VAMPIR approach.
\par
The differences in the binding energies of {\sl real} MONSTER and {\sl real}
VAMPIR are negligibly small. The even spin {\sl real} MONSTER yrast states,
$2^+$, $4^+$ and $6^+$, are on average only 56~keV more bound. The
$0^+$ state does not mix with the projected two-quasiparticle states
of the same symmetry in the {\sl real} case, too.
\subsection{Odd-odd nuclei}
\label{oddodd}
In order to study the quality of the {\sl complex} MONSTER approach for
applications to odd-odd nuclei, we calculated the energy spectra of
$^{20}_{\ 9}$F, $^{22}_{11}$Na, $^{24}_{11}$Na, $^{26}_{13}$Al,
$^{28}_{13}$Al and $^{30}_{15}$P. For each nucleus the {\sl complex} MONSTER 
configuration space was built on the HFB transformation, which yielded the 
energetically lowest VAMPIR solution in just this nucleus, no matter which 
spin it corresponded to. The spectra have also been calculated with {\sl real} 
MONSTER. In this case one has to take the HFB transformation of a neighbouring
even-even nucleus to construct the MONSTER configuration space. Naturally, one
has several possibilities~: for an odd-odd nucleus with Z protons and N neutrons
one can take the even-even neighbours with the proton and neutron numbers
(Z-1,N-1), (Z-1,N+1), (Z+1,N-1), and (Z+1,N+1). It is a priori not clear,
which choice may be the best. Therefore we looked in each neighbour
for the HFB transformation, which belongs to the $0^+$ ground state
VAMPIR solution and used each for a subsequent {\sl real} MONSTER calculation.
\par
The energies obtained by both MONSTER approaches have been compared
to the ones of a full shell model configuration mixing calculation.
To get a quantitative measure, for each nucleus as already for 
the even-even nuclei, the average deviation of the energies of the yrast
states with spins $0^+$ to $5^+$ from the shell model energies has been
calculated. The resulting deviations are shown in table~2, again in keV and in
percent of the shell model ground state energy. Besides the energy
the table displays the spin of the shell model ground state and, for 
{\sl complex} MONSTER the spin of the VAMPIR solution which was used. For
{\sl real} MONSTER always the result for that spectrum which yielded the best
agreement with the shell model is listed in the table. The corresponding
neighbouring nucleus is indicated. The {\sl real} MONSTER spectra for
one odd-odd nucleus, obtained from the HFB transformations of the various
possible neighbours, differ often quite much. So, e.g., for $^{22}$Na
the average deviation of the yrast states $0^+$ to $5^+$ is 4.85\% if the
HFB transformation for the $0^+$ VAMPIR solution of $^{22}$Ne is used,
and 1.47\% for the one of $^{20}$Ne. 
\par
Table~2 also shows the minimum and maximum number of shell model
configurations per spin and the number per spin averaged for spins
$0^+$ to $5^+$. Since protons and neutrons are mixed by the {\sl complex} HFB
transformation, the same configurations are used for the calculation of
even-even and odd-odd nuclei in a {\sl complex} MONSTER calculation. Thus the
number of configurations contributing to one spin given in section
(\ref{eveneven}) is valid here, too. For the {\sl real} MONSTER with no
proton-neutron mixing the situation is different compared to the even-even
case. For the description of odd-odd nuclei in $sd$ shell there are
14 configurations available for the $0^+$ states, 39 for $1^+$, 57 for $2^+$,
67 for $3^+$, 71 for $4^+$ and always 72 for the states with spins
$5^+, \ldots 13^+$. These numbers are comparable in size to the ones of the
even-even case.
\par
As already for the even-even nuclei, also for the odd-odd ones an increase 
of the average deviation with the number of shell model states per spin is
observed. For $^{20}$F the whole shell model spectrum is exactly reproduced by
{\sl complex} MONSTER using any {\sl complex} HFB transformation $F$. This is
to be expected, since $^{20}$F belongs to the same isospin multiplet like
$^{20}$Ne. $^{20}$Ne encloses states of all isospins $T\! =\! 0, 1, 2$,
$^{20}$F only the ones with isospins $T\! =\! 1$ and $2$. This provides
another stringent test for the numerics. The largest deviation relative to
the shell model ground state energy was found for $^{24}$Na~: 2.21\%.
$^{24}$Na is a typical example for which one should better choose different
HFB transformations for states with different spins, instead of using one
transformation for all states. If on uses the HFB transformation of the $0^+$
VAMPIR solution for the low spin states $0^+$ to $3^+$ and the transformation
F($4^+$) for the higher spins, the average deviation of the yrast states
$0^+$ to $5^+$ can be reduced to 995~keV or 1.26\% of the shell model ground
state energy. The possibility to choose different HFB transformations for
different spins should always be considered, since we found that the
{\sl complex} MONSTER energies are quite sensitive on the underlying HFB
transformation in contrast to the {\sl real} case \cite{BEN95a}.
\par
The average deviations of the {\sl real} MONSTER are always bigger than
for the {\sl complex} approach, apart from $^{24}$Na, where both are of
similar size. But this could be improved by using several HFB
transformations in the {\sl complex} case. As far as the higher excited 
states are concerned, the {\sl complex} MONSTER is always superior to the
{\sl real} one, since it is able to describe many more states.
\par
Compared to the even-even case the deviations obtained by {\sl complex}
MONSTER are somewhat bigger for each mass number. It is known that the
structure of odd-odd nuclei is generally more complicated. More types of
couplings contribute significantly. This seems to be reflected in our results.
But also for the odd-odd case the exact spectra are over all rather
well approximated by the MONSTER approach.
\par
In fig.~3 the energy spectra of $^{22}$Na calculated with the {\sl complex}
and the {\sl real} MONSTER are plotted up to spin 7 and an energy of $-52$~MeV.
For comparison the lowest shell model energies for each spin are presented.
In the {\sl real} case, MONSTER is built on the HFB transformation determined
for the $0^+$ VAMPIR solution in $^{20}$Ne. This gave the best agreement with
the shell model. The most bound {\sl complex} VAMPIR solution in this
nucleus is the one for the $1^+$ yrast state. {\sl Complex} MONSTER on top of
the corresponding HFB transformation corrects then the sequence of the yrast
states to the one of the exact calculation, where the ground state has spin
$3^+$. The largest deviation of a {\sl complex} MONSTER yrast state energy
from the shell model value occurs for the $0^+$ yrast state~: 362~keV,
the smallest for $7^+$~: 83~keV. {\sl Complex} MONSTER shows
also for the higher lying states an excellent agreement with the energies
of the exact shell model and yields a significant improvement on the
{\sl real} MONSTER.
\par
Finally, as an example for the heavier odd-odd nuclei, fig.~4 displays the
calculated energies of $^{30}$P. In this figure two {\sl real} MONSTER
calculations are presented, the ones yielding the best and the worst agreement
with the result of the exact diagonalization. The former has been got by using
the HFB transformation for the $0^+$ VAMPIR solution in $^{28}$Si, the latter
by the HFB transformation for the $0^+$ VAMPIR solution in $^{32}$S.
The lowest {\sl complex} VAMPIR solution in $^{30}$P was found to be the
one with spin $0^+$. Its HFB transformation was used in the {\sl complex}
calculation. The {\sl complex} and the {\sl real} MONSTER results are
plotted up to $\sim$11~MeV above the shell model ground state for spins
$0^+$ to $5^+$. For each spin the lowest energies of the exact
diagonalization are given. {\sl Complex} MONSTER reproduces the general
trends of the spectrum rather well. The largest deviation of the yrast 
state energies is found for spin $2^+$ and amounts to 1579~keV or 1.01\% of
the shell model ground state energy, the smallest,
1157~keV or 0.74\%, for spin $0^+$. The {\sl complex} calculation shows a
clear improvement on both {\sl real} ones, where much less states are
described.
\subsection{Odd mass nuclei}
\label{oddmass}
As described in section (\ref{restrictions}), in the case of odd mass nuclei
the {\sl complex} MONSTER spectrum has to be built on a HFB transformation
determined for a neighbouring even mass nucleus. We proceeded here in
the following way. We calculated the shell model spectra of several of those 
odd mass nuclei, which are neighbouring the even mass nuclei discussed 
in sections (\ref{eveneven}) and (\ref{oddodd}). Then we used the HFB
transformations, which we had already determined, to built up the MONSTER
one-quasiparticle configuration spaces. Accordingly, for the mass parameter
of the Chung-Wildenthal force, we took always the mass of the even mass
nucleus of which the HFB transformation was used. The energy spectra of
$^{21}_{10}$Ne, $^{23}_{11}$Na, $^{25}_{12}$Mg, $^{27}_{13}$Al and
$^{29}_{14}$Si have been calculated. For $^{21}$Ne the HFB transformations
of the deepest VAMPIR solutions in $^{22}$Ne and $^{22}$Na have been used. The
other odd mass nuclei with mass number A have been calculated always in four
different ways~: using the HFB transformation for the deepest VAMPIR solution
of the even-even neighbour with mass A$-$1, the one of the even-even neighbour
with mass A$+$1 and the ones of the odd-odd neighbours with A$-$1 and A$+$1.
Thus, e.g., $^{29}$Si has been calculated using the mean fields of $^{28}$Si,
$^{30}$Si, $^{28}$Al and $^{30}$P. In the case of {\sl real} MONSTER only
the mean fields of the two neighbouring even-even nuclei are available.
Both have been used for the {\sl real} calculations.
\par
The MONSTER configuration space for the calculation of odd mass nuclei is
much smaller than the one for the even mass nuclei, because there
are much less one-quasiparticle than two-quasiparticle excitations.
For {\sl complex} MONSTER in the $sd$ shell there are 12 configurations
available for states with spin $I\! =\! 1/2$, 20 for $I\! =\! 3/2$ and
always 24 for states with higher spins. The {\sl real} MONSTER provides even
only 3 configurations for $I\! =\! 1/2$ states, 5 for $I\! =\! 3/2$, and always
6 for all higher spin states.
\par
Let us first turn our attention to $^{21}$Ne. It is the lightest odd mass
nucleus which we have calculated. For this nucleus there are in total 1935
shell model configurations. More precisely, there are 199 $I\! =\! 1/2$,
341 $I\! =\! 3/2$, 400 $I\! =\! 5/2$, 368 $I\! =\! 7/2$, 287 $I\! =\! 9/2$,
183 $I\! =\! 11/2$, 100 $I\! =\! 13/2$, 41 $I\! =\! 15/2$, 14 $I\! =\! 17/2$,
and 2 $I\! =\! 19/2$ states. Fig.~5 shows for each spin the lowest energies
resulting from the complete shell model diagonalization up to spin
$I\! =\! 11/2$. In addition the results of two {\sl complex} MONSTER
calculations are displayed up to an energy of $-38$~MeV. One calculation
was done by using the mean field of the $0^+$ ground state for $^{22}$Ne, the
other is built on the mean field of the deepest VAMPIR solution for $^{22}$Na,
i.e., the $1^+$ solution. In fig.~5 also the result of a {\sl real} MONSTER
calculation is plotted for comparison. It is built on the HFB transformation
determined for the $0^+$ ground state of $^{22}$Ne by a {\sl real} VAMPIR
calculation. The obtained agreement between both {\sl complex} MONSTER
results and the shell model is very good. For the calculation built on
$^{22}$Ne the average deviation of the yrast state energies from the exact
ones is 490~keV. That is 1.03\% of the shell model ground state energy,
which is $-47.753$~MeV. The largest deviation occurs for the $11/2$ state
and amounts to 664~keV or 1.39\% of the shell model ground state energy,
the smallest for spin $5/2$, and is 256~keV or 0.54\%. The calculation using
the mean field of $^{22}$Na gives an average deviation of 463~keV or 0.97\%
measured relative to the shell model ground state energy. The largest
deviation, 653~keV or 1.37\%, was found for the $1/2$ state, the smallest,
308~keV or 0.64\%, for the $7/2$ yrast state. Thus, the mean field of the
$1^+$ yrast state of $^{22}$Na is on average slightly better suited for
the description of the yrast states of $^{21}$Ne than that of the $0^+$
ground state of $^{22}$Ne. Except for the spin $1/2$ and $5/2$ yrast states,
it yields deeper bound states. Also the excited states obtained by this
calculation display the bigger binding energies. The {\sl real} MONSTER
gives for the yrast state energies an average deviation of 576~keV or
1.21\% from the shell model ones. The $1/2$ state shows the largest
deviation~: 704~keV or 1.47\%, the $9/2$ state the smallest~:
467~keV or 0.98\%. For most of the yrast states both {\sl complex} MONSTER
calculations yield more binding than the {\sl real} ones, as it is expected. 
However, e.g., for the $9/2$ yrast state the {\sl real} MONSTER yielded
the deepest energy. The {\sl complex} results, built on the mean fields of
$^{22}$Na and $^{22}$Ne, are higher by 16~keV and 114~keV, respectively.
This indicates that in the {\sl complex} case, the mean field determined
for a low spin may not be well suited for the description of high spin states.
\par
As example for an heavier odd mass nucleus, we present the energy
spectra of $^{29}$Si obtained by using the mean fields of the mass 28 systems.
The number of shell model states in this nucleus is huge compared to the
number of MONSTER configurations available. In total this nucleus has
80112 shell model states. 5638 of these states have spin $I\! =\! 1/2$,
10176 $I\! =\! 3/2$, 12877 $I\! =\! 5/2$, 13450 $I\! =\! 7/2$,
12240 $I\! =\! 9/2$, 9835 $I\! =\! 11/2$, 7053 $I\! =\! 13/2$,
4469 $I\! =\! 15/2$, 2502 $I\! =\! 17/2$, 1197 $I\! =\! 19/2$,
485 $I\! =\! 21/2$, 152 $I\! =\! 23/2$, 35 $I\! =\! 25/2$,
and 3 are spin $I\! =\! 27/2$ states. For each spin up to spin $9/2$ the
lowest energies resulting from the exact shell model calculation are shown
in fig.~6. The obtained MONSTER spectra are plotted up to an energy of
$\sim$ 10~MeV above the shell model ground state energy of $-147.020$~MeV.
One {\sl complex} MONSTER spectrum is obtained by using the HFB transformation
of the VAMPIR solution for the $0^+$ ground state in $^{28}$Si, the other
uses the transformation of the $0^+$ VAMPIR solution in $^{28}$Al. The
calculation on top of the mean field of the even-even nucleus shows an
average deviation of the yrast state energies from the exact ones of 1373~keV
or 0.93\% in terms of percentage of the shell model ground state energy.
The yrast state with spin $9/2$ shows the largest deviation, 1575~keV or
1.07\%, the $7/2$ yrast state the smallest, 1291~keV or 0.88\%.
The average deviation obtained by using the mean field of the odd-odd nucleus
is somewhat bigger, namely 1912~keV or 1.30\%. The largest and smallest
deviations, which occur, are 2172~keV or 1.48\% and 1607~keV or 1.09\%,
respectively. The general trend of the shell model spectrum is reproduced
by both {\sl complex} calculations. In fig.~6 in addition also a {\sl real}
MONSTER spectrum is shown. It was obtained by using the HFB transformation
of the $0^+$ VAMPIR solution in $^{28}$Si. This spectrum yielded a better
agreement with the shell model result than the {\sl real} calculation on top
of the mean field for $^{30}$Si. For the displayed {\sl real} MONSTER spectrum
the average deviation of the yrast state energies from the exact ones is
1973~keV or 1.34\%. The largest deviation occurs for the $1/2$ yrast state,
2286~keV or 1.55\%, the smallest for $7/2$, 1640~keV or 1.12\%. The 
{\sl complex} MONSTER calculation on top of the mean field of
$^{28}$Si shows a clear improvement over this best {\sl real} result.
\par
For all the considered odd mass nuclei, apart from $^{21}$Ne, we found that
by using the HFB transformation of a neighbouring even-even nucleus, a better
agreement between the {\sl complex} MONSTER result and the shell model could
be achieved than by using the mean field of an odd-odd neighbour. In general
the odd mass spectra on top of a mean field of a neighbouring even-even
nucleus approximated the shell model results well (more details can be found
in \cite{BEN95c}). The only exceptions were the spectrum of $^{23}$Na obtained
by using the HFB transformation of the $0^+$ VAMPIR solution for $^{22}$Ne
and the spectrum of $^{25}$Mg with the transformation for $^{26}$Mg. These two
calculations gave a rather poor agreement with the exact result. In both cases
it could be much improved by using the HFB transformation of $^{24}$Mg.
Thus, in the odd mass case the choice of the neighbour can be crucial.
One should always try several ones to find the mean field, of which the
structure is best suited for the odd mass nucleus under consideration.
\par
Finally we investigated for the odd mass case the dependence of the
{\sl complex} MONSTER energies on the spin of the mean field determined by
the preceding VAMPIR calculation. For $^{21}$Ne we built always the full
MONSTER spectrum on each of the yrast solutions of $^{22}$Ne obtained by
{\sl complex} VAMPIR. The same was done for {\sl real} MONSTER. Naturally
here only mean fields for even spin states are available.
\par
Fig.~7 shows a typical result of this investigation. The energies of the
five lowest $3/2$ states of $^{21}$Ne are plotted in dependence on
the spin used in the preceding VAMPIR calculation for $^{22}$Ne.
For comparison, the leftmost column presents the energies of the
three lowest $3/2$ shell model states. Solid lines denote the results of the
{\sl complex} MONSTER, dotted ones those of the {\sl real} counterpart.
It can be seen that {\sl real} MONSTER displays only a weak dependence on
the spin of the underlying HFB transformations. For the {\sl complex}
MONSTER this is also true if the VAMPIR transformations of the low spin
mean fields, $0^+$, $2^+$ and $4^+$, are considered. The mean fields of the
higher spin states are insufficient for the description of the $3/2$ states.
The poor description of the $3/2$ states using the VAMPIR transformations
of the low odd spins, $1^+$, $3^+$ and $5^+$, is due to the fact that
those yrast states are dominated by structures which are missing in a
{\sl complex} VAMPIR calculation \cite{BEN95a}.
\par
The same behaviour as for the $3/2$ states was also found for most of the
spins of $^{21}$Ne. Only for the very high spins, where the number of MONSTER
configuration is of similar size as the number of shell model states, the
dependence becomes small. When the number of MONSTER configurations is bigger
than the number of shell model states, the resulting energies are independent
of the underlying transformation.
\par
This investigation confirms what we have found already for the even mass case
\cite{BEN95a}. The {\sl complex} HFB transformation shows a much
stronger dependence on the spin of the underlying VAMPIR calculation than the
{\sl real} one. It means that the structure of the {\sl complex} mean field
differs more from spin to spin. The choice of the spin of the underlying
VAMPIR calculation is important in the {\sl complex} description of odd mass
nuclei, too.
\section{Summary}
\label{summary}
A recently developed nuclear structure approach, the {\sl complex} MONSTER,
has been studied. The approach is the newest member of a group of models,
which are based on variational methods. This group can be divided into two
subgroups, one consisting of the VAMPIR models, the other of the MONSTER
approaches. Both are designed for large scale nuclear structure calculations.
\par
The models of the VAMPIR family use symmetry-projected HFB quasiparticle
vacua as test wavefunctions. The underlying HFB transformations are
determined by variation. By construction, these models can only be used to
describe the lowest few states of a certain symmetry. In the first
numerically realized VAMPIR models the HFB transformations were rather
restricted~: time-reversal invariance and axial symmetry were required,
parity and proton-neutron mixing were neglected and only {\sl real} HFB
transformation coefficients were allowed. In the newest models most
restrictions are removed and essentially {\sl complex} HFB transformations
are allowed. Only time reversal and axiality are kept.
\par
In the MONSTER approaches the nuclear wavefunctions are expanded around
a VAMPIR solution for the ground or an yrast state. The spectrum of excited
states is obtained by diagonalizing the Hamiltonian in the space of the
VAMPIR solution and all symmetry-projected two-quasiparticle configurations
with respect to it. These models are therefore suited for problems,
where a complete set of excitations with respect to a particular transition
operator is needed. Till lately the MONSTER approach was restricted
to the use of {\sl real} VAMPIR solutions. In a recent work it was generalized.
Now {\sl complex} HFB transformations can be used to build up the configuration
space. Then more nucleon correlations are considered already in the mean field
and the configuration space becomes much larger. Consequently a more
detailed description of nuclear spectra is expected by this {\sl complex}
MONSTER approach. Furthermore one can use HFB transformations determined
for the particular system under consideration not only for even-even systems,
as in the {\sl real} approach, but also for odd-odd systems. Only for
odd mass nuclei, which are described in the one-quasiparticle approximation,
one still has to use the mean field of a neighbouring nucleus, which now may
be even-even or odd-odd. To get insight into the ability of this {\sl complex}
approach a systematic investigation was performed in this work.
\par
We presented first results of an application of the {\sl complex} MONSTER
approach to several nuclei in $sd$ shell. This basis, which is rather tiny
for a MONSTER calculation, has been chosen to enable a comparison with complete
shell model configuration mixing calculations. The investigated even-even
and odd-odd nuclei have masses between A$=$20 and 30, the odd mass nuclei
reach from A$=$21 to 29. Thus, quite light systems with only a few valence
nucleons are considered as well as heavier systems, up to $^{28}$Si with
the biggest number of shell model configurations in $sd$ shell.
\par
For the even-even nuclei the {\sl complex} MONSTER yielded an excellent
agreement with the exact shell model approach. The largest mean deviation
of the yrast state energies from the shell model result, averaged over
the states with spins $0$ to $5$, was found for $^{26}$Mg and amounts only
to 1.43\% of its shell model ground states energy. Concerning the odd-odd
systems it was demonstrated that {\sl complex} MONSTER reproduced the
shell model spectrum of $^{20}$F exactly. $^{24}$Na showed the largest
average deviation of the energies of the yrast states with spins $0$ to $5$
from the shell model ones, namely 2.21\% of the ground state energy.
The average deviations of the odd-odd systems were slightly bigger than the
ones of the even-even systems with the same mass number. But in general
the {\sl complex} MONSTER yielded a very good approximation to the shell
model result also for odd-odd nuclei. For both even-even and odd-odd nuclei
a clear improvement on the previous {\sl real} approach was proved.
\par
For odd mass nuclei it was found that the description depends much on the
underlying mean field. By choosing a suitable mean field a good agreement
with the shell model could be achieved for all investigated nuclei. As
examples the results of $^{21}$Ne and $^{29}$Si were presented. For $^{21}$Ne,
using the HFB transformation of the $1^+$ VAMPIR solution in $^{22}$Na, the
deviation of the yrast states energies, averaged over spins $1/2$ to $11/2$,
was 0.97\% of its shell model ground state energy. For $^{29}$Si,
with the mean field determined for the ground state of $^{28}$Si, the average
deviation of the yrast states energies for spins $1/2$ to $9/2$ amounted to
0.93\% relative to the ground state energy. Again the {\sl complex} 
approach achieved better agreement with the shell model than the
{\sl real} one.
\par
Already the previous {\sl real} MONSTER has been applied successfully to the
description of nuclear structure phenomena in various mass regions, e.g.,~in
the mass 130 region \cite{HAM85} \cite{HAM86}. The present investigation has
clearly shown that the new {\sl complex} approach is even superior. So we
expect that the {\sl complex} MONSTER approach will develop as a more
powerful tool for nuclear structure studies in large model spaces.
\par
Up to now only the energies calculated by the different models have been
compared. It will be interesting to look also for other observables,
like transition strengths up to high excitation energies, and to analyze
the wavefunctions. This is planned for the future. Finally we would like to
mention that the description of nuclear wavefunctions by VAMPIR and
MONSTER could be improved by removing all restrictions imposed on the
HFB transformation. The development of such approaches allowing the most
general HFB transformations is in progress, too.
\\[2.5\parskip]
We thank Prof.~Dr.~Herbert M\"uther for performing the shell model
calculations with the Glasgow code. Furthermore this work was partly
supported by the Gra\-du\-ier\-ten\-kol\-leg ``Struktur und Wechselwirkung
von Hadronen und Kernen'' (DFG, Mu 705/3).
\newpage
\newpage
\section*{Figure Captions}
\begin{description}
\item[Fig.~1]
Comparison of the energy spectra of $^{28}$Si for spins 0 to 6 as obtained
by three different approaches~: the shell model configuration mixing approach
(SCM), the {\sl complex} MONSTER (CM) and the {\sl real} MONSTER (RM).
The energy is given relative to the $^{16}$O core. For each spin only
the lowest few shell model states are presented. The MONSTER results are 
displayed up to excitation energies of $\sim$~14~MeV above the shell model
ground state. The {\sl complex} and the {\sl real} MONSTER calculation has
been built on the corresponding {\sl complex} and {\sl real} VAMPIR solution
for the $0^+$ ground state.
\item[Fig.~2]
The energies of the yrast states with spin $0^+$ to $6^+$ in $^{28}$Si as
obtained by various approaches~: the shell model (SCM), the {\sl complex} 
MONSTER (CM) (based on the $0^+$ solution of {\sl complex} VAMPIR), the
{\sl complex} VAMPIR (CV), the {\sl real} MONSTER (RM) (on top of the
$0^+$ solution of {\sl real} VAMPIR), and the {\sl real} VAMPIR (RV).
In the last approach only the even spin states are accessible.
Spin and parity are indicated on the l.h.s.~of each level. The energy is
given relative to the $^{16}$O core.
\item[Fig.~3]
Same as Fig.~1, but for the odd-odd nucleus $^{22}$Na and up to spin 7.
Each MONSTER spectrum is shown up to 7.5~MeV above the shell model ground
state. The {\sl complex} MONSTER was built on the {\sl complex} VAMPIR
solution obtained for the $1^+$ yrast state, the {\sl real} MONSTER was based
on the {\sl real} VAMPIR solution for the $0^+$ ground state of $^{20}$Ne.
\item[Fig.~4]
Same as Fig.~1, but for the odd-odd nucleus $^{30}$P and up to spin 5.
The results of two {\sl real} MONSTER calculations are displayed. Each
MONSTER spectrum is shown up to 11~MeV above the shell model ground state.
The {\sl complex} MONSTER was built on the {\sl complex} VAMPIR solution
obtained for the $0^+$ yrast state. One {\sl real} MONSTER calculation has
been based on the VAMPIR transformation for the $0^+$ ground state of
$^{28}$Si, the other on the one of the $0^+$ ground state of $^{32}$S.
\item[Fig.~5]
Same as Fig.~1, but for the odd mass nucleus $^{21}$Ne and for spins
1/2 to 11/2. The results of two {\sl complex} MONSTER calculations are
displayed. Each MONSTER spectrum is shown up to $-$38~MeV. One {\sl complex}
MONSTER calculation has been based on the {\sl complex} VAMPIR transformation
for the $0^+$ ground state of $^{22}$Ne, the other was using the one for the
$1^+$ yrast state of $^{22}$Na. The {\sl real} MONSTER calculation was built
on the {\sl real} VAMPIR solution for the $0^+$ ground state of $^{22}$Ne.
\item[Fig.~6]
Same as Fig.~1, but for the odd mass nucleus $^{29}$Si and for spins
1/2 to 9/2. The results of two {\sl complex} MONSTER calculations are
displayed. Each MONSTER spectrum is shown up to 10~MeV above the shell
model ground state. One {\sl complex} MONSTER calculation has been based
on the {\sl complex} VAMPIR transformation for the $0^+$ ground state of
$^{28}$Si, the other was using the one for the $0^+$ yrast state of $^{28}$Al.
The {\sl real} MONSTER calculation was built on the {\sl real} VAMPIR
solution for the $0^+$ ground state of $^{28}$Si.
\item[Fig.~7]
The five lowest $3/2^+$ MONSTER states of $^{21}$Ne are plotted versus
the spin of the yrast state in $^{22}$Ne, of which the VAMPIR transformation
was used in each calculation. Solid lines refer to the {\sl complex} MONSTER,
dotted lines to the  {\sl real} MONSTER results. In the latter case only
transformations for even spin values are available. For comparison the
lowest three $3/2^+$ shell model (SCM) energies are also given. The energy is 
given relative to the $^{16}$O core.
\end{description}
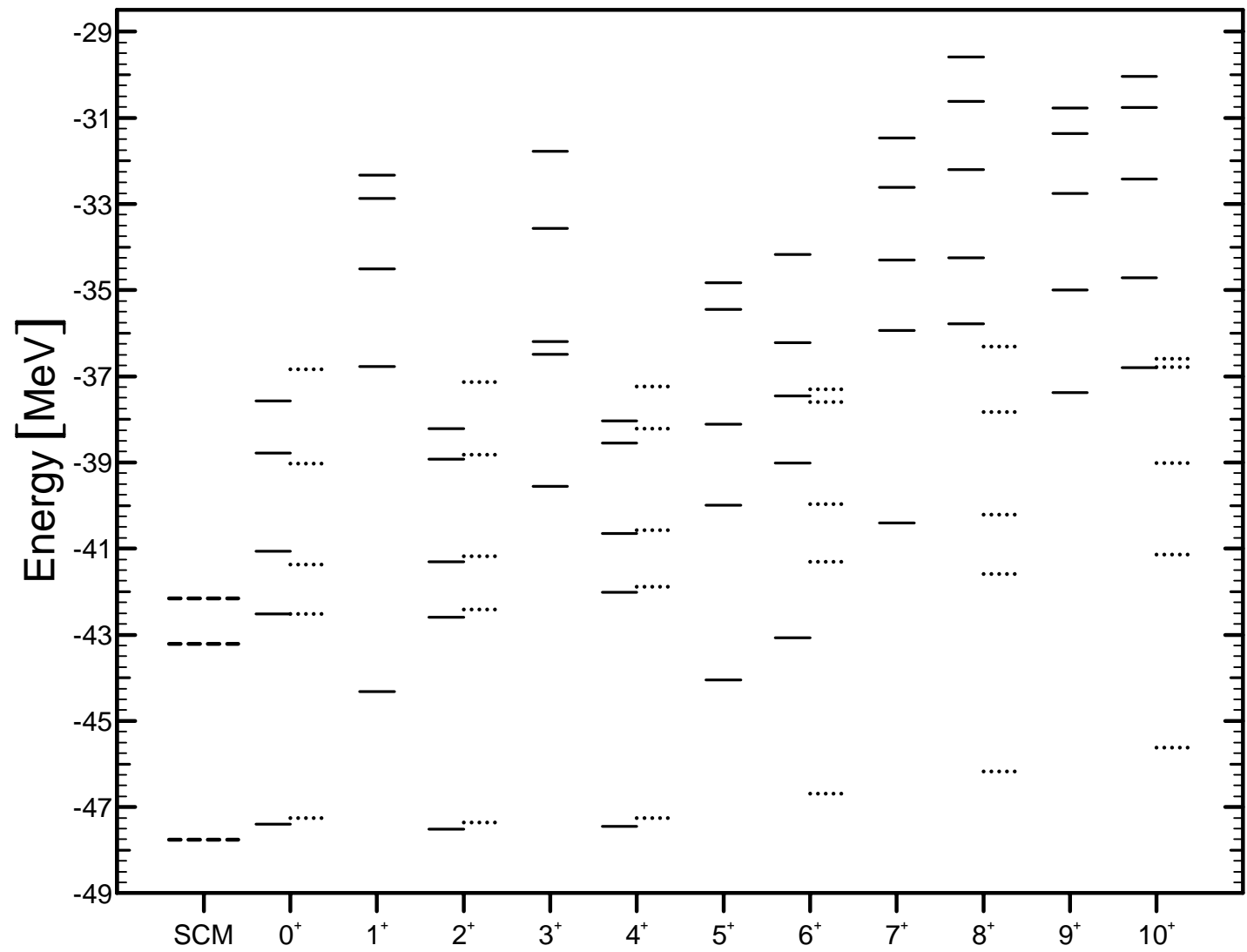
\end{document}